\documentclass[sort&compress]{aastex701}
\setcitestyle{square,numbers,comma,sort&compress}

\usepackage{amsmath,amsfonts,amssymb}
\usepackage{multirow}
 \usepackage{graphicx}
 \usepackage{tocloft}
 \usepackage[dvipsnames]{xcolor}

 \usepackage{longtable}
 \usepackage{lineno}

\newcommand{\kms}{km~s$^{-1}$}

\begin{document}

\nolinenumbers 

\title{Revealing Cosmic Ecosystems with the Hubble Space Telescope in 2030s and Beyond}

\author[0000-0002-2724-8298, sname=Borthakur, gname=Sanchayeeta]{Sanchayeeta Borthakur}
\affiliation{(Affiliations listed in the Appendix)}
\email[show]{sanch@asu.edu}

\author[0009-0001-5959-9105]{Tanmay Singh}
\affiliation{(Affiliations listed in the Appendix)}
\email{tsingh65@asu.edu}

\author[0000-0003-3520-6503]{David French}
\affiliation{(Affiliations listed in the Appendix)}
\email{a@b}

\author[0000-0003-3520-6503]{Yakov Faerman}
\affiliation{(Affiliations listed in the Appendix)}
\email{a@b}

\author[0000-0001-6248-1864]{Kate Rubin}
\affiliation{(Affiliations listed in the Appendix)}
\email{a@b}

\author[0000-0001-5530-2872]{Brad Koplitz}
\affiliation{(Affiliations listed in the Appendix)}
\email{a@b}

\author[0000-0002-3120-7173]{Rongmon Bordoloi}
\affiliation{(Affiliations listed in the Appendix)}
\email{a@b}

\author[0000-0003-4237-3553]{Frances H. Cashman}
\affiliation{(Affiliations listed in the Appendix)}
\email{a@b}

\author[0000-0001-8587-218X]{Matthew J. Hayes}
\affiliation{(Affiliations listed in the Appendix)}
\email{matthew.hayes@astro.su.se}   

\author[0000-0003-4158-5116]{Yong Zheng}
\affiliation{(Affiliations listed in the Appendix)}
\email{zhengy14@rpi.edu}

 \author[0000-0002-1979-2197]{Joseph N. Burchett}
\affiliation{(Affiliations listed in the Appendix)}
\email{a@b} 

\author[0000-0003-4877-9116]{Jane C. Charlton}
\affiliation{(Affiliations listed in the Appendix)}
\email{a@b}

\author[0000-0001-8813-4182]{Hsiao-Wen Chen}
\affiliation{(Affiliations listed in the Appendix)}
\email{a@b}

\author[0000-0003-0724-4115]{Andrew J. Fox}
\affiliation{(Affiliations listed in the Appendix)}
\email{a@b}

\author[0000-0002-6137-0422]{Yucheng Guo}
\affiliation{(Affiliations listed in the Appendix)}
\email{a@b}

\author[0000-0001-6670-6370]{Timothy M. Heckman}
\affiliation{(Affiliations listed in the Appendix)}
\email{a@b}

\author[0000-0002-2591-3792]{Christopher J. Howk}
\affiliation{(Affiliations listed in the Appendix)}
\email{a@b}

\author[0000-0001-9487-8583]{Sean D. Johnson}
\affiliation{(Affiliations listed in the Appendix)}
\email{a@b}


\author[0000-0003-1362-9302]{Glenn G. Kacprzak}
\affiliation{(Affiliations listed in the Appendix)}
\email{a@b}

\author[0000-0002-2587-2847]{Varsha P. Kulkarni}
\affiliation{(Affiliations listed in the Appendix)}
\email{a@b}

\author[0000-0001-9158-0829]{Nicolas Lehner}
\affiliation{(Affiliations listed in the Appendix)}
\email{a@b}

\author[0000-0003-3938-8762]{Sowgat Muzahid}
\affiliation{(Affiliations listed in the Appendix)}
\email{a@b}

\author[0000-0002-4430-8846]{Namrata Roy}
\affiliation{(Affiliations listed in the Appendix)}
\email{a@b}

\author[0000-0002-3193-1196]{Evan Scannapieco}
\affiliation{(Affiliations listed in the Appendix)}
\email{a@b}

\author[0000-0002-0355-0134]{Jessica K. Werk}
\affiliation{(Affiliations listed in the Appendix)}
\email{jwerk@uw.edu}

\begin{abstract}

Ultraviolet spectroscopy with the Hubble Space Telescope (HST) provides the most direct and sensitive probe of the disk–circumgalactic medium (CGM) interface at radii of 20 kpc, where galaxies exchange gas, metals, and energy with their surroundings. Many of the key diagnostics of the multiphase circumgalactic medium -- including H~I, O~VI, C~II–IV, Si II–IV, N~V, Ne~VIII, and other metal transitions -- lie in the ultraviolet and are inaccessible from the ground, making HST the only observatory capable of making the required observations. By measuring the physical (column density, density), chemical (metallicity, ionization structure), and kinematical properties of the gas at the disk-CGM interface, UV absorption-line spectroscopy reveals how galaxies acquire fresh fuel, recycle enriched material, and drive feedback into their halos. When combined with spectroscopic characterization of the host galaxy's stellar populations and the feedback they generate (outflow velocity, mass loading), we will establish a direct understanding of how stellar populations enable circulation of gas and metals through the galactic ecosystem. HST's ultraviolet (UV) spectroscopic capability provides the only comprehensive observational pathways for uncovering the physical drivers that regulate galaxy growth and evolution in the low-redshift Universe.

\end{abstract}

\section{The Big Questions of the next Decade}

One of the central themes identified by the Astro2020 Decadal Survey is the study of the {\it Cosmic Ecosystem} --- the interconnected processes that govern how galaxies acquire, process, and expel baryons over cosmic time. Despite major advances in our understanding of the diffuse circumgalacic medium (CGM) surrounding galaxies \citep{chen1998,Kulkarni2005,Prochaska2005,steidel2010,Tumlinson11,tripp2011,Churchill2013,werk14,Savage2003,borthakur13,borthakur15,heckman17,fox2015,Hayes2016,Prochaska2017,rudie2019,bordoloi2014,Kacprzak15,chen2020,Peroux2020,Peroux_Howk2020,Cashman2021,sameer2021,Kulkarni2022,bordoloi2025,zheng2024,pessa2024,Nateghi2024a,Nateghi2024b,Kacprzak2025,Dutta2025}, the physical mechanisms regulating gas exchange between galaxies and their CGM remain poorly constrained. 
Quantifying these gas flows and characterizing their energetics, pathways, masses, thermal structure, and spatial variation represent critical missing pieces that are essential for establishing a comprehensive framework for the galactic ecosystem.

\begin{figure}[t]
\begin{center}
\includegraphics[trim= 0mm 0mm 2mm 0mm, clip=true, width=6.0in]{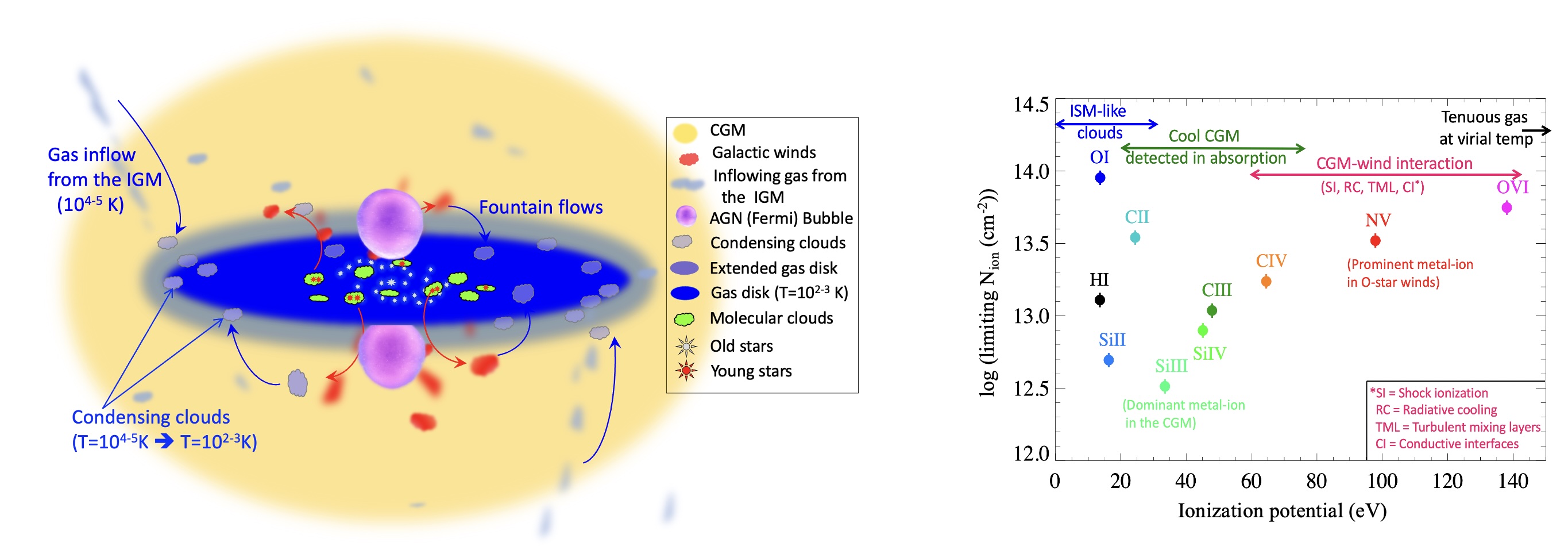}
\end{center}
\vspace{-0.45cm}
\caption{{\em Left:} Schematic of the cosmic ecosystem -- showing gas flows that connect the disk to the circumgalactic medium. The disk–CGM interface is the primary nexus of baryon cycling in galaxies, where gas accretion, galactic outflows, cooling, mixing, and feedback-driven fountain flows interact to regulate the exchange of mass, metals, and energy between galaxies and their surrounding halos. {\em Right:} The range of ionization potential explored via transitions in the rest-frame FUV region of $\rm 1032 \le \lambda (\AA) \le 1550 $ as a function of limiting column density for absorption-line observations at S/N $\sim$10 (for $\rm W_{\sigma} \le 70~m\AA$). The availability of transitions covering the entire range of energies (ionization potentials) is crucial in distinguishing between various physical models, including the non-equilibrium processes. }
\vspace{-0.25cm}
\label{fig:BaryonCycle_Cartoon}
\end{figure}

The disk--circumgalactic medium (CGM) interface, a transition zone within $\sim20-30$ kpc from a galaxy's star-forming disk, is a uniquely important regime where the majority of baryonic flow processes either originate or terminate \citep{borthakur2025,fraternali2017,fraternali2008,howk2018,lehner2026,Charlton1996}. It is within this transition region that the CGM can be directly linked to the interstellar medium (ISM) and the stellar component, thereby providing a complete view of how gas accretes onto galaxies, cycles through the ISM, and is ultimately expelled back into the halo \citep{rubin2014,bordoloi2011,bouche2013,Schroetter2019}. Figure~\ref{fig:BaryonCycle_Cartoon} presents a schematic view of a Milky Way-like galaxy highlighting the diverse physical processes operating at the disk--CGM interface.

\subsection{Scope of Discovery}

The disk--CGM interface remains a major discovery space that has yet to be systematically explored, primarily due to the limited availability of sensitive ultraviolet (UV) spectroscopic facilities. Extending the lifetime of the Hubble Space Telescope would enable transformative, long-duration investments in QSO absorption-line spectroscopy of the disk--CGM interface, combined with down-the-barrel spectroscopy of galaxy disks. Such a program would provide an unprecedented opportunity to directly connect the physical conditions of the CGM, interstellar medium (ISM), and stellar populations within a unified framework of baryon cycling and galaxy evolution.
Such observations would enable us to:
\vspace{-0.3cm}
\begin{itemize}
\item[1.] Quantify gas accretion rates and identify the dominant pathways through which baryons enter galaxy disks;
\vspace{-0.3cm}
\item[2.] Trace feedback-driven outflows from parsec-scale energy injection sites into the extended CGM;
\vspace{-0.3cm}
\item[3.] Characterize the stellar populations and energetic sources responsible for regulating the gaseous ecosystem; and
\vspace{-0.3cm}
\item[4.] Search for relic AGN-driven structures analogous to the Milky Way’s Fermi Bubbles and determine their impact on the surrounding CGM.
\end{itemize}

Processes like gas accretion, galactic fountains, stellar feedback, turbulence, mixing, and AGN-driven outflows may all operate simultaneously, producing a dynamically coupled ecosystem \citep{Martin1999,Strickland2009,fox_dave2017,Somerville2015,Hopkins2014,nelson2019}. 
Although these structures may overlap in projection, each process imprints distinct signatures on the gas content, kinematics, ionization structure, metallicity, and thermal state, enabling them to be disentangled observationally.

However, to do so we requires access to a broad suite of ultraviolet transitions spanning a wide range of ionization potentials and physical conditions. The rest-frame far-UV contains the highest density of diagnostic transitions capable of tracing the multiphase gas responsible for baryon and energy exchange across the disk--CGM interface (see the right panel of Figure~\ref{fig:BaryonCycle_Cartoon}). Table~\ref{tab:lines_phase} highlights several of the key UV transitions 

\begin{deluxetable*}{c l l}[!b]
\tablewidth{5pt}
\tablecaption{Dominant lines and their origin and importance.}
    \tablehead{Ion & Lines (\AA) & Tracer/Importance } 
    \startdata
        H~I & Lyman series (1215--912) & Neutral gas signatures $\equiv$ traces the bulk gas content. \\
        H$_2$$^*$& 1150--912&  Molecular gas signature. \\
        O{\sc i} & 1302  & Metallicity tracer for neutral gas.\\   
        O{\sc vi}$^*$ & 1032,1038  & Dominant coolant tracing multiple energetic processes and a strong coronal-line doublet.\\
        C{\sc ii} &   1304 & A strong metal lines enable detection of low column density gas.\\
        C{\sc iii}$^*$ & 977 & One of the strongest metal-lines in diffuse media. \\      
        C{\sc iv}$^\dagger$ & 1548, 1551 & One of the three coronal transitions accessible for low-z galaxies. \\  
        N{\sc v}$^\dagger$ & 1239, 1243 & One of the three coronal transitions accessible for low-z galaxies. \\          
        Si{\sc ii}  &  1190,1193,1260,1304,1527 & Single species with five lines of different intrinsic strength to counter saturation effects.\\
        Si{\sc iii} &   1206 & Most commonly detected metal-line detected in the CGM.\\
        Si{\sc iv}  &   1394,1403 & Together with Si{\sc ii} and Si{\sc iii} enables ionization correction.\\
        S{\sc ii}   & 1250,1253,1259 &  Low-depletion species used for metallicity measurements.\\
        Al{\sc ii}$^\dagger$& 1670& Traces denser ISM-like gas and enable depletion studies. \\
    \enddata
\label{tab:lines_phase}
\tablecomments{$^*$ Lines are only accessible when redshifted. $^\dagger$ Need access to near-UV grating for slightly redhsifted systems }
\end{deluxetable*}

\begin{figure}[t]
\begin{center}
\includegraphics[trim= 0mm 0mm 2mm 0mm, clip=true, width=5.0in]{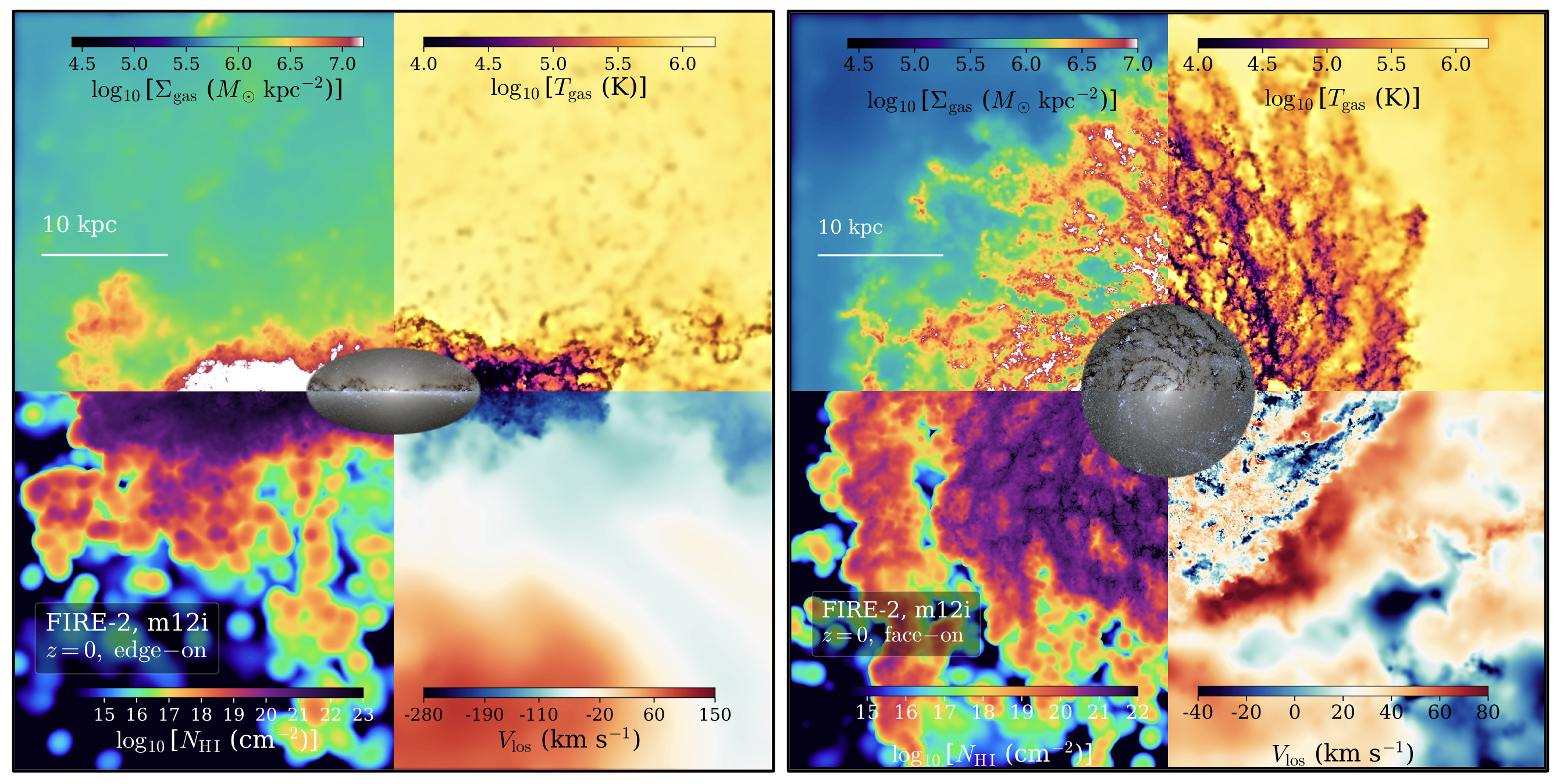}
\end{center}
\vspace{-0.45cm}
\caption{Edge-on (left) and face-on (right) views of a Milky Way-mass FIRE-2 zoom-in galaxy \texttt{m12i} at $z=0$ \citep{Wetzel2016,Wetzel2023}, with $M_\star = 6.3\times10^{10}\,M_\odot$ and ${\rm SFR}=9.2\,M_\odot\,{\rm yr}^{-1}$, shown across a $60\,{\rm kpc}\times60\,{\rm kpc}$ region centered on the galaxy. Each view combines four projected gas diagnostics: gas surface density in the upper-left quadrant, gas temperature in the upper-right quadrant, neutral hydrogen column density in the lower-left quadrant, and line-of-sight velocity in the lower-right quadrant. The central stellar mock image provides the disk reference: the upper elliptical region shows dust-attenuated stellar light, while the lower elliptical region shows the dust-free view. Together, the projections highlight the complex and multiphase structure of the disk--halo interface, including dense neutral gas, warm/hot gas, and coherent velocity structure around the galaxy.}
\vspace{-0.25cm}
\label{fig:simulations}
\end{figure}

Figure~\ref{fig:simulations} presents edge-on and face-on views of a simulated galaxy from the FIRE-2 simulations, illustrating the highly structured and non-axisymmetric nature of the gaseous medium. Rather than exhibiting smooth radial profiles, the gas distribution displays complex filamentary and clumpy structures produced by ongoing accretion, turbulence, cooling, and feedback-driven outflows, highlighting the dynamical processes that regulate galaxy evolution.

\section{Observational Design} \label{sec:observations}

Constraining how gas flows into, through, and out of galaxies requires observations that simultaneously resolve gas kinematics at $\sim$20~\kms\ and measure accurate column densities across multiple ionization states, enabling metallicity, density, and temperature constraints on the multiphase medium. This demands both a statistical sample of single QSO--galaxy sightlines and a smaller set of galaxies probed by multiple background QSOs within $\sim$25~kpc of the disk. Together, these observations would deliver the first topological view of the disk--CGM interface, overcome stochastic pencil-beam variations, and establish a robust framework for mapping baryon cycling in galaxies.

A critical next step is to directly connect the physical conditions of the disk--CGM interface to the energetic sources embedded within galaxy disks that drive feedback processes. Down-the-barrel rest-frame far-UV spectroscopy provides a uniquely powerful diagnostic of both stellar populations and the surrounding ISM \citep{Heckman2015, Martin2012,Chisholm2017}. Stellar wind features constrain the ages, energetics, and radiation fields of young stellar populations, while low- and intermediate-ionization absorption lines trace the kinematics, column densities, and mass-loading of galactic outflows \citep{Leitherer2011,Schwartz2006,James2014}. Together, these observations enable a direct link between star formation, energy injection, and the multiphase gaseous structures that regulate matter and energy exchange between galaxies and their halos \citep[e.g.][]{Martin2019}.

\subsection{Targets and Requirements} \label{sec:observations}

Target selection for disk--CGM interface studies must prioritize nearby galaxies with UV-bright background QSOs probing a broad range of impact parameters, azimuthal angles, galaxy masses, morphologies, and star-formation properties. In particular, a successful program requires sightlines extending from the edge of the optical disk out to the inner 10\% of the CGM ($\lesssim$ 20--30~kpc), where inflows, galactic fountains, turbulence, and feedback-driven outflows are expected to imprint strong signatures in the gaseous medium \citep{Oppenheimer2010,Ford2014,Hopkins2018,Corlies2020,Peroux2024,Stern2024,Oren2026}.

Equally important is the identification of a subset of galaxies intersected by multiple QSO sightlines, enabling tomographic constraints on the distribution, coherence scale, and covering fraction of multiphase gas structures \citep[e.g.][]{lehner2026, bowen2016}. Combining these absorption-line measurements with ancillary H~I \citep{Westmeier2018,Das2024}, optical IFU \citep{Burchett2021, Guo2023,Nielsen2024,Zhang2024}, optical and sub-mm (Sunyaev-Zel'dovich effect) spectral stacking \citep{Zhang2018b,Das2025}, and multiwavelength imaging \citep{Hayes2016,Rupke2019,Ha2023,pessa2024} data will provide the environmental and energetic context necessary to interpret the baryon cycle.

The primary limitation in constructing such samples is the low surface density of sufficiently bright ultraviolet background QSOs. Current Cosmic Origins Spectrograph (COS) \citep{Green2012} studies are largely restricted to the brightest UV sources, which significantly limits the number of accessible galaxy--QSO pairs and biases observations toward sparse spatial sampling of halos. This limitation becomes particularly severe at small impact parameters, where the probability of finding a bright background QSO intersecting the disk--CGM interface is intrinsically low. As a result, most of the CGM studies so far focused beyond the disk-CGM interface with the exception of Milky Way and local group galaxies \citep{fox2006, mishra2024,Barger2013,Richter2017}. 

Restricting observations to only the brightest QSOs strongly biases samples toward sparse sightline coverage and limits our ability to probe the critical disk--CGM transition region at small impact parameters. Extending the operational lifetime of HST would enable the community to pursue deeper integrations on substantially fainter QSOs, dramatically increasing the available target density on the sky. Access to fainter background sources is essential for constructing statistically significant samples, obtaining multiple sightlines through individual halos, and probing rare but astrophysically crucial regions near galaxy disks. 

\begin{figure}[t]
\includegraphics[trim= 0mm 0mm 2mm 0mm, clip=true, width=7in]
{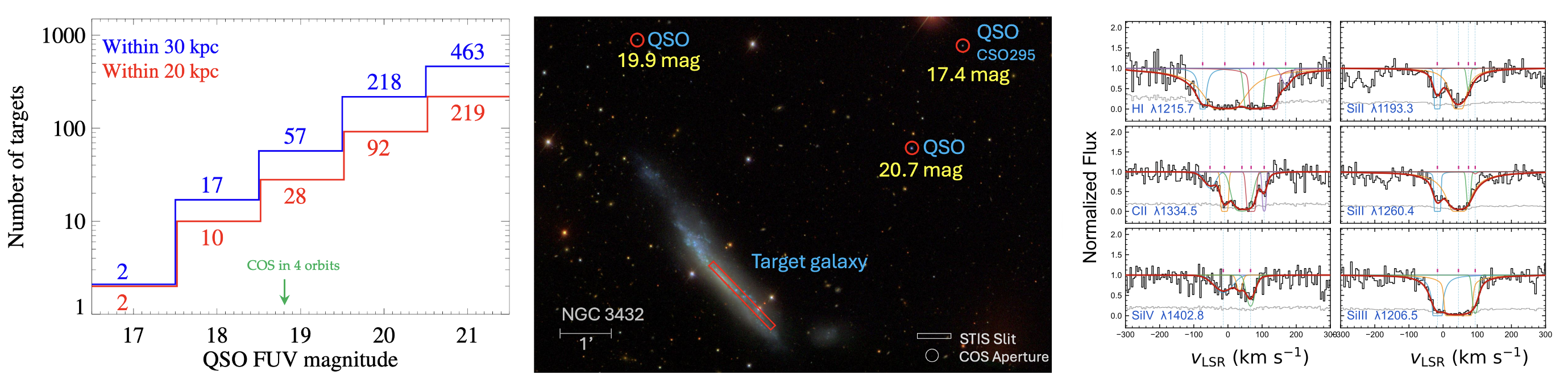}
\caption{{\em Left:} Number of galaxy pairs as a function of UV magnitude of the background QSO. The redshift range of the foreground galaxies was chosen to be between 0.002$<$z$<$0.1  (recession velocity $>$600 $\rm km~s^{-1}$). The red and the blue lines indicate QSOs within 20 and 30~kpc of the foreground galaxy, respectively. The pairs were obtained by cross-matching GALEX UV sources with the spectroscopic galaxy catalog from the Sloan Digital Sky Survey. Although the numbers might change if using other galaxy catalogs due to differences in their sky coverages, spectroscopic completeness, and other biases, the general trend is expected to be the same. The sensitivity of the state-of-the-art spectrograph, COS, is shown as a green arrow. The COS can achieve an S/N = 10 for a limiting FUV magnitude of 18.8 in four orbits at 1250 $\rm \AA$ where its sensitivity is optimal. {\em Center:} We show a COS/STIS setup around NGC 3432 that covers three QSO sightlines and the stellar body of the galaxy. The red circle shows the position of QSOs and the red rectangle represents the STIS slit. The annotations are not to scale \citep[adapted from][]{borthakur2025}. {\em Right:} COS G130M spectra showing complex gas structure at the position of the top-right QSO (CSO~295) \citep{french2020}.}
\label{fig:Observational_setup}
\end{figure}

Targeting QSOs as faint as GALEX FUV $\sim$19--21 mag is essential to achieve the sightline density needed to probe galaxies with a range of stellar mass at low impact parameters, and varied azimuthal angles, as well as to utilize rare systems with multiple background sightlines through a single halo. As shown in Figure~\ref{fig:Observational_setup}, extending to these fainter UV sources increases the number of QSO--galaxy pairs by a factor of $\sim$3, transforming studies of the disk--CGM interface from sparse pencil-beam measurements into a statistically robust mapping of baryon cycling.

To robustly constrain the physical conditions of the multiphase gas, the observations must achieve signal-to-noise ratios of $\sim$8--10 per resolution element across the suite of diagnostic transitions listed in Table~\ref{tab:lines_phase}. Reaching this sensitivity is critical for accurately measuring weak and moderately saturated absorption features spanning a broad range of ionization states, enabling reliable determinations of gas kinematics, column densities, ionization structure, metallicity, and thermal conditions. Such data are required to disentangle overlapping inflow, outflow, and recycling signatures and to perform detailed ionization modeling of the gas phases that regulate baryon exchange between galaxy disks and the CGM.

\subsection{Programmatic Needs} \label{sec:observations}
A major opportunity for enabling these systematic studies lies in expanding sustained access to COS and Space Telescope Imaging Spectrogragh (STIS) onboard the HST.
A comprehensive program designed to map the disk--CGM interface across a statistically meaningful sample of galaxies would require hundreds of orbits ($<$500-1000 orbits) to achieve the necessary sensitivity (S/N$\sim$10), spectral resolution (R$\sim$20,000), and spatial sampling. Such an investment is difficult to realize within the framework of standard observing allocations and will require a coordinated community effort supported by dedicated long-term access to HST resources. 
The need for spectral resolution and sensitivity makes this an HST only program and not suitable other planned missions (e.g., UVEX).
Establishing a {\em community-driven} legacy program would enable the large, homogeneous datasets needed to transform UV absorption-line studies of the disk--CGM interface from isolated case studies into a predictive framework for understanding baryon cycling and galaxy evolution.

The instrumental capabilities and operational requirements for this program are fully consistent with the current performance of the HST. The background QSOs targeted for absorption-line spectroscopy are unresolved point sources, enabling straightforward target acquisition and unperturbed by mild pointing inaccuracy with existing COS and STIS observing modes \citep{Mas-Hesse2003, Revalski2018}. Consequently, the proposed observations impose no new operational or technical constraints beyond standard HST operations, making this program immediately feasible within the current observatory framework.

\section{Preparatory Science for the Habitable Worlds Observatory} \label{sec:HWO}

The HST can serve as a direct pathfinder for the Habitable Worlds Observatory by exploiting its existing far-ultraviolet spectroscopic capabilities to establish the observational foundation for next-generation studies of the disk--CGM interface. Through QSO absorption-line spectroscopy, HST can empirically determine the physical scales, ultraviolet diagnostics, sensitivity thresholds, spectral resolution requirements, and target-selection strategies necessary for mapping the disk-CGM interface to trace the baryon cycle.

By extending HST/COS observations to fainter, but substantially more numerous, background QSOs, it becomes possible to construct the first statistically significant UV absorption-line census of the disk--CGM interface in nearby galaxies. Such observations would move the field beyond isolated pencil-beam measurements toward a systematic characterization of how gas and metals accrete onto galaxies, circulate through the ISM, and are expelled into the CGM through feedback processes. The resulting datasets would provide direct constraints on the multiphase structure, kinematics, metallicity, and topology of the gaseous ecosystem regulating galaxy evolution.

At present, many of the predicted instrument requirements and survey strategies for HWO are driven primarily by theoretical simulations. However, different simulations adopt substantially different prescriptions for star formation, feedback, cooling, turbulence, and baryon transport, leading to large variations in the predicted observables. A number of these key observables remain accessible with current HST capabilities, provided sufficiently deep integrations can be obtained. The principal limitation is therefore not technical feasibility, but the availability of the large observing-time allocations required to reach the sensitivity necessary for statistically meaningful samples of faint background QSOs. A dedicated long-term HST legacy effort would therefore provide the empirical framework needed to anchor HWO science requirements in observations rather than simulations alone.

\subsection{Synergies with existing and upcoming observatories}

Such a large HST legacy program will work synergistically with next-generation observations to advance our understanding of gas flows in and around galaxies. These include H~I 21 cm surveys mapping extended disks, diffuse extraplanar gas, and high-velocity clouds; optical IFU spectroscopy tracing stars and CGM gas in emission; fast radio burst surveys probing electron densities; and X-ray observations revealing the hotter gas phases. 

\section{Recommendations} \label{sec:recommendation}

We recommend a Hubble Space Telescope legacy program for far-ultraviolet QSO absorption-line spectroscopy of the disk--CGM interface in nearby galaxies. HST uniquely enables medium- and high-resolution far-UV spectroscopy of the diffuse multiphase gas that regulates galactic ecosystems, directly addressing the Astro2020 priority of understanding the {\it Cosmic Ecosystem}.

The program should obtain deep COS and STIS spectra of UV-bright background QSOs probing galaxies over a broad range of masses and environments, with emphasis on sightlines within $\sim$10--50~kpc of galaxy disks where inflows and outflows interact. Simultaneous observations of the host galaxy should provide strong constraints on feedback.

Extending observations to fainter QSOs ($\rm FUV\sim19$--20 mag) is essential for achieving the sightline density needed to map the gaseous ecosystem at small impact parameters. Spectra with S/N $\sim$8--10 would constrain gas kinematics, column densities, metallicities, and ionization structure.

\newpage

\bibliographystyle{unsrt}
\bibliography{sample701}

\appendix

\section{Author Affiliations}

\begin{tabular}{ll}
Sanchayeeta Borthakur & School of Earth \& Space Exploration, 
Arizona State University (ASU), 781 Terrace Mall\\
& Tempe, AZ 85287, USA\\
Tanmay Singh  & Arizona State University, 781 Terrace Mall, Tempe, AZ 85287, USA\\
David French & Space Telescope Science Institute, 3700 San Martin Drive, Baltimore, MD 21218, USA\\
Yakov Faerman & School of Physics and Astronomy, Tel Aviv University, Ramat Aviv 69978, Israel \\
Kate Rubin & Department of Astronomy, San Diego State University, San Diego, CA 92182, USA   \\
Brad Koplitz  & Arizona State University, 781 Terrace Mall, Tempe, AZ 85287, USA\\
& Tempe, AZ 85287, USA \\
Frances H. Cashmen & Department of Physics, Presbyterian College, 503 South Broad Street Clinton, SC 29325\\
Yong Zheng & Department of Physics, Applied Physics and Astronomy, Rensselaer Polytechnic \\
& Institute, Troy, NY 12180, USA\\
Matthew J. Hayes & Stockholm University, Department of Astronomy and Oskar Klein Centre for  \\
& Cosmoparticle Physics, AlbaNova University Centre, SE-10691, Stockholm, Sweden \\
Rongmon Bordoloi & Department of Physics, North Carolina State University, Raleigh, NC 27695, USA\\
Joseph N. Burchett  & Department of Astronomy, New Mexico State University, PO Box 30001, USA\\
Jane C. Charlton  & Department of Astronomy and Astrophysics, The Pennsylvania State University,\\
& State College, PA 16801, USA\\
Hsiao-Wen Chen &Department of Astronomy \& Astrophysics, The University of Chicago, 5640 S Ellis Ave., \\
&Chicago, IL 60637, USA\\
Christopher Churchill & Department of Astronomy, New Mexico State University, PO Box 30001, USA\\
Andrew J. Fox & AURA for ESA, Space Telescope Science Institute, 3700 San Martin Drive, \\
& Baltimore, MD 21218, USA\\
Timothy M. Heckman &  Department of Physics and Astronomy, The Johns Hopkins University, \\
 &Baltimore, MD 21218, USA \\
Christopher J. Howk & Department of Physics, University of Notre Dame, Notre Dame, IN 46556, USA\\

Yucheng Guo & Arizona State University, 781 Terrace Mall, Tempe, AZ 85287, USA\\
Sean D. Johnson &Department of Astronomy, University of Michigan, 1085 S. University, Ann Arbor, \\
&MI 48109, USA\\
Glenn G. Kacprzak& Centre for Astrophysics and Supercomputing, Swinburne University of Technology,\\
& Hawthorn, Victoria 3122, Australia; ARC Centre of Excellence for All Sky \\
& Astrophysics in 3 Dimensions (ASTRO 3D), Australia\\
Varsha P. Kulkarni & Department of Physics and Astronomy, University of South Carolina, Columbia, \\
& SC 29208, USA\\
Nicolas Lehner & Department of Physics and Astronomy, University of Notre Dame, Notre Dame, \\
&IN 46556, USA\\
Sowgat Muzahid & Inter-University Centre for Astronomy \& Astrophysics, Post Bag 04, Pune 411007,India\\
Namrata Roy & Arizona State University, 781 Terrace Mall, Tempe, AZ 85287, USA \\
Evan Scannapieco & School of Earth \& Space Exploration, 
Arizona State University, 781 Terrace Mall \\
& Tempe, AZ 85287, USA \\
Jessica K. Werk & Department of Astronomy, University of Washington, Seattle, WA 98195, USA \\
\end{tabular}

\end{document}